\documentclass{cjaa}

\usepackage{graphicx}                   %for PS/EPS graphics inclusion, new

\begin{document}

\title{Seven-frequency VLBI observations of the GHz-Peaked-Spectrum source OQ 208}

   \volnopage{Vol.0 (200x) No.0, 000--000}      %%preserved for Editor. DOn't remove!
   \setcounter{page}{1}          %%starting page, preserved for Editor. DOn't remove!

\author{Wen-Feng Luo            \inst{1,3}
        \and Jun Yang       \inst{2,3}
        \and Lang Cui       \inst{1,3}
        \and Xiang Liu      \inst{1} \mailto{}
        \and Zhi-Qiang Shen \inst{2,4}
        }

\offprints{Wen-Feng Luo}
\institute{%
          National Astronomical Observatories/Urumqi Observatory, Urumqi  830011, China\\
          \email{liux@ms.xjb.ac.cn}
          \and Shanghai Astronomical Observatory, Chinese Academy of Sciences, Shanghai 200030, China
          \and Graduate University of the Chinese Academy of Sciences, Beijing 100049, China
          \and Joint Institute for Galaxy and Cosmology, SHAO and USTC, China}

   \date{Received~~  ~~month day; accepted~~ ~~month day}

\abstract{%
We present images of quasi-simultaneous VLBI  observations of the
GHz-Peaked-Spectrum  radio source OQ 208 with the Very Long Baseline
Array at 1.4, 1.7, 2.3, 5.0, 8.4, 15.4 GHz and the European VLBI
Network at 6.7 GHz. The low frequency (1.4, 1.7 and 2.3 GHz)
observations reveal a weak and extended steep-spectrum component at
about 30 mas away in the position angle of $- 110^\circ$ which may
be a remnant emission. The radio structure of OQ 208 consists of two
mini-lobes at 5.0, 6.7, 8.4 and 15.4 GHz. Our spectral analysis
further confirms that the southwest
 lobe undergoes free-free absorption  and finds that the free-free absorption
is stronger in the inner region. By fitting the 8.4 GHz images from
1994 to 2005, we obtain a separation speed of 0.031 $\pm$ 0.006 mas
yr$^{-1}$ between the two mini-lobes. This indicates a jet proper
motion of 0.105 $\pm$ 0.020 $c$ and a kinematic age of 219 $\pm$ 42
yr for the radio source.
}

\authorrunning{W.-F. Luo et al.}
\titlerunning{Seven-frequency VLBI observations of OQ 208}

\maketitle

\section{introduction}
GHz-peaked-spectrum (GPS) sources, a sub-class of powerful radio
sources ($L_\mathrm{radio} \approx 10^{45}$ erg s$^{-1}$), lie in
the Narrow Line Region with characteristics of a steep rising
spectrum at low frequencies. Synchrotron self-absorption (SSA) or
free-free absorption (FFA) has been proposed for the inverted
spectra (O'Dea \cite{ode98}). Both mechanisms are supported by some
observations (e.g. Yang et al. \cite{yan05}). The previous VLBI
observations indicate that most of GPS galaxies exhibit either a
compact double (CD) or a compact symmetric object (CSO) morphology
(e.g. Stanghellini et al. \cite{sta01}).

The radio source OQ 208 (B1404+286, J1407+2827; J2000.0: 14h7m0.394
+ 28d27'14.69"),  is one of the closest ($z$ = 0.077) GPS sources
with a spectral peak at 4.9 GHz (Dallacasa et al. \cite{dal00}). It
is not resolved with the Very Large Array (VLA). VLBI observations
(Fey et al. \cite{fey96}; Stanghellini et al. \cite{sta97}) revealed
double mini-lobes separated by $\sim$ 7 mas in NE-SW direction.
Based on the multi-epoch VLBA observations at 2.3/8.4 GHz,
Stanghellini et al. (\cite{sta97,sta00}) identified OQ 208 as a CSO,
and determined a separation speed of 0.033 $\pm$ 0.013 mas yr$^{-1}$
between the two lobes and a kinematic age of 204 $\pm$ 81 yr.

The host galaxy of OQ 208 has a brighter core ($m_r$=14.6) and broad
recombination lines and is classified as a Seyfert 1 galaxy (de
Grijp et al. \cite{gri92}) or broad line radio galaxy (Marziani et
al. \cite{mar93}). The optical image displays a low-brightness tail
in the north direction toward east and indicates the presence of
companions in the galactic envelope (Stanghellini et al.
\cite{sta93}). The X-ray observation (Guainazzi et al. \cite{gua04})
discovered that OQ 208 is a Compton-thick active galactic nucleus
(AGN).

The compactness of OQ 208 and its proximity to us make it a good
candidate for the studies of absorption mechanisms, proper motions
of jet components, and how this class of sources develops with
high-resolution VLBI observation.

In this paper, we present results of the quasi-simultaneous VLBI
observations at seven frequencies (1.4, 1.7, 2.3, 5.0, 6.7, 8.4 and
15.4 GHz ). A new weak component at $\sim$ 30 mas in the  position
angle of $-110^\circ$ is detected and the multi-frequency radio
images and components' spectra have been obtained. The proper motion
between the two mini-lobes is estimated based on our and previous
8.4 GHz VLBI data. We adopt the cosmological model with $H_0 = 70$
km s$^{-1} $Mpc$^{-1}$,  $\Omega_m = 0.3$, $\Omega_\Lambda = 0.7$
and define $S_\nu \propto \nu^{\alpha}$ throughout the paper.

\section{Observations and data reduction}
\begin{table}[b]
\centering  \label{tab1} \caption{The related parameters of the
observations and the images in Fig. 1. Col. (1), serial number of
the panels. Col. (2), observation frequency in GHz. Col. (3),
observation date. Col. (4), array name. Col. (5), recording rate in
Mbps. Col. (6), program ID. Col. (7), peak flux density in mJy/beam.
Col. (8), the lowest contour level (3$\sigma$) in mJy/beam.
Col.(9-11), size and position angle of the restoring beam in mas and
degree.} \scriptsize\label{tab1}
\begin{tabular}{crccccrrrrr}
 \hline \hline
   & Freq.  &Date          & Array & Rate    &Program ID & $S_\mathrm{peak}$  &  L.C.   & Maj.     & Min.   & P.A.  \\
   & (GHz)  &              &       & Mbps    &           &  (mJy/b)           & (mJy/b) & (mas)    & (mas)  & ($^\circ)$\\
\hline
a & 1.438  & May 1, 2005 & VLBA  & 256     & BW0080     &     714.7          &    1.3  &   16.10  &  5.66  &  8.09    \\
b & 1.667  & May 3, 2005 & VLBA  & 128     & BY0020     &     926.6          &    2.1  &   8.24   &  4.42  &  5.94    \\
c & 2.270  & May 3, 2005 & VLBA  & 128     & BY0020     &     1290.1         &    1.0  &   9.81   &  8.94  &  42.30   \\
d & 2.270  & May 3, 2005 & VLBA  & 128     & BY0020     &     1147.4         &    3.6  &   5.19   &  3.48  &  $-$3.06 \\
e & 4.987  & May 3, 2005 & VLBA  & 128     & BY0020     &     1524.9         &    3.8  &   2.53   &  1.33  &  $-$1.30 \\
f & 6.668  & Nov. 11, 2004& EVN   & 256     & EN003B     &     978.1          &    11.7 &   5.88   &  0.67  &  77.80    \\
g & 8.420  & May 3, 2005 & VLBA  & 128     & BY0020     &     985.7          &    3.2  &   1.31   &  0.96  &  1.18    \\
h & 15.365 & May 3, 2005 & VLBA  & 128     & BY0020     &     346.3          &    2.0  &   0.80   &  0.54  &  0.75   \\
\hline
%\hline
\end{tabular}
\end{table}

Table 1 lists some basic information of the VLBI observations. In
the VLBI observations, the radio source OQ 208 was used as a fringe
finder. Our VLBI observations (BY0020) were carried out at 1.7,
2.3/8.4, 5.0 and 15.4 GHz with the Very Long Baseline Array (VLBA)
on May 3, 2005. OQ 208 was observed for 4 scans at 1.7 and 5.0 GHz,
6 scans at 2.3/8.4 and 15.4 GHz.  Each scan lasts 4 minutes. Such
repeated snap-shot mode of multiple-scans gives a good uv-coverage.
The observations were made using left circular polarization, four
8-MHz channels and 2 bit sampling. The data were correlated at
Socorro with 2-second integration time, 16 channels and uniform
weight.

The data at 1.4 and 6.7 GHz are analyzed in order to obtain the
better spectral coverage. The 1.4-GHz data are provided by Wrobel
who observed OQ 208 with an 11-minute scan and 256 Mbps recording
rate with the VLBA. The 6.7-GHz data  are from the EVN data archive
(PI: Bartkiewicz). The EVN array consists of 9 stations (Cm, Jb, Ef,
Mc, On, Tr, Nt, Hh and Wb). In this observation, OQ 208 was observed
for 13 minutes.

The $a$-$priori$ calibrations were done with the NRAO Astronomical
Imaging Processing Software (AIPS) package (Cotton \cite{cot95}). At
each frequency, the amplitude calibration was performed using the
system temperature measurements and antenna gain and the phase
solutions were derived using the global fringe fitting with a
2-minute solution interval and a point-source model. After checking
the solutions, we applied the solutions to the data, averaged all
the channels in each IF and split the multi-source data into the
single source data.

The self-calibration and imaging process were performed using the
DIFMAP package (Shepherd et al. \cite{she94}). The overall amplitude
self-calibration was not performed until the clean models had the
amplitude close or equal to that of the short-baseline visibility.
The gain correction for each antenna is a small factor (within 1
$\pm$ 0.15). We also fitted the calibrated visibility data to the
Gaussian models using the MODELFIT program. The approximate errors
of the integrated flux density and position are calculated using the
formula given by Lobanov
\footnote{http://www.radionet-eu.org/wikiattach/SchoolOrganisationPages/attachments/lobanov.pdf}.
The model-fitting results are reported in Table 2.

\begin{figure}
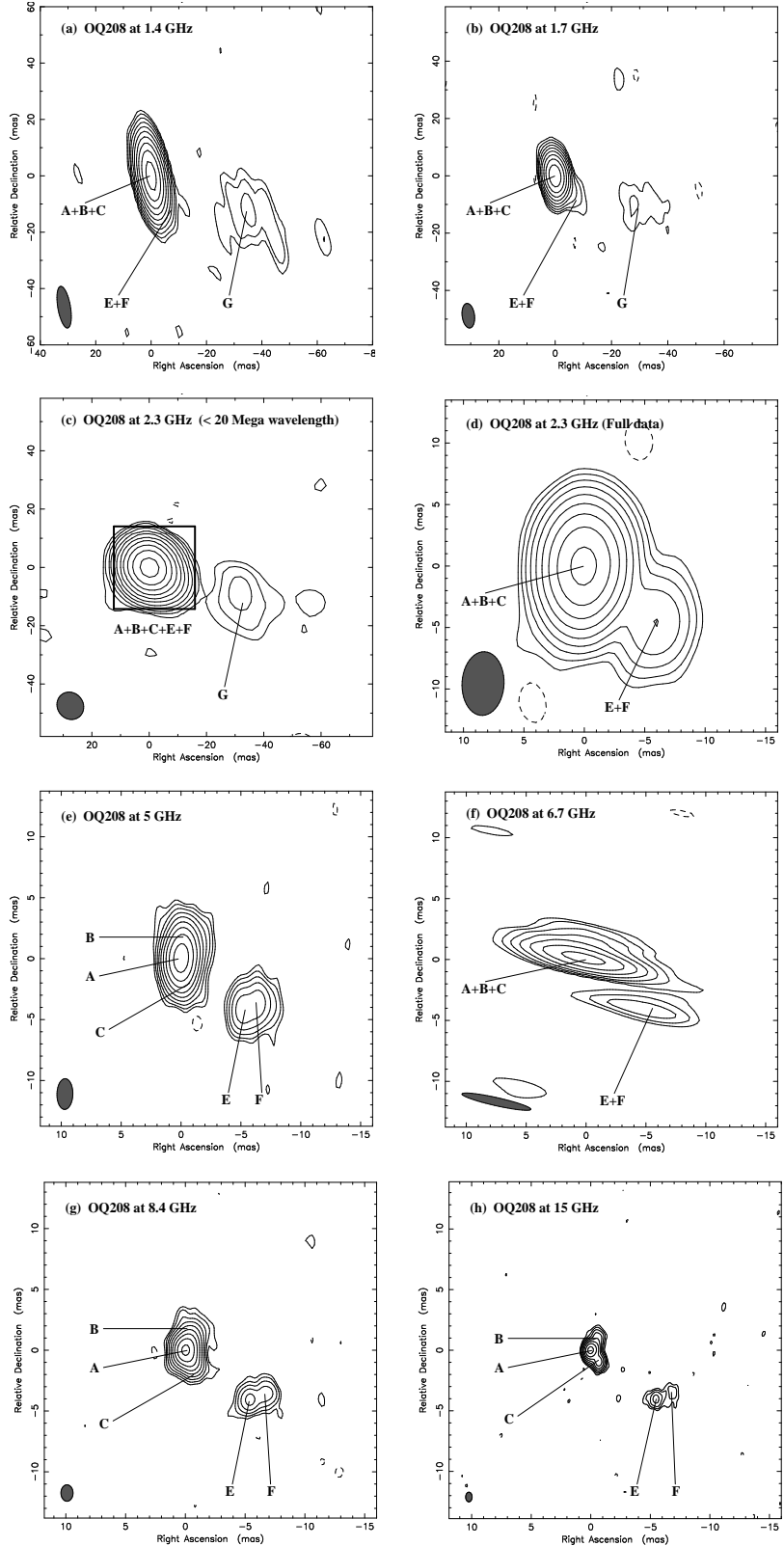

\centering \label{fig1} \caption{The intensity images of OQ 208 from
high-resolution VLBI observations at 1.4, 1.7, 2.3, 5.0, 6.7, 8.4
and 15.4 GHz. The lowest contour level is three times the off-source
r.m.s noise level. The contour levels increase by a factor of 2. The
restoring beam is shown as an ellipse in the lower-left corner. The
image parameters are listed in Table \ref{tab1}. The fitting
parameters of the marked Gaussian components are listed in Table
\ref{tab2}. }
\includegraphics[width=0.35\textwidth, height=0.35\textwidth]{OQ208_1.4GHz.eps}
\hspace{0.02\textwidth}
\includegraphics[width=0.35\textwidth, height=0.35\textwidth]{OQ208_1.7GHz.eps}
\vspace{0.02\textwidth}\\
\includegraphics[width=0.35\textwidth, height=0.35\textwidth]{OQ208_2.3GHz_s.eps}
\hspace{0.02\textwidth}
\includegraphics[width=0.35\textwidth, height=0.35\textwidth]{OQ208_2.3GHz.eps}
\vspace{0.02\textwidth}\\
\includegraphics[width=0.35\textwidth, height=0.35\textwidth]{OQ208_5.0GHz.eps}
\hspace{0.02\textwidth}
\includegraphics[width=0.35\textwidth, height=0.35\textwidth]{OQ208_6.7GHz.eps}
\vspace{0.02\textwidth}\\
\includegraphics[width=0.35\textwidth, height=0.35\textwidth]{OQ208_8.4GHz.eps}
\hspace{0.02\textwidth}
\includegraphics[width=0.35\textwidth, height=0.35\textwidth]{OQ208_15.0GHz.eps}
\\
\label{fig1}
\end{figure}

\section{results}
\subsection{Radio morphology}

The final images are displayed in Fig. 1. In each image, the lowest
contour level is three times the off-source rms noise level. The
contour levels increase by a factor of 2. All the images are
restored using the uniform weight. The restoring beam is shown as an
ellipse in the lower-left corner. The peak flux density, the lowest
contour level, the size, and the position angle of the restoring
beam are listed in Table~1. The model-fitting parameters of the
marked components are listed in Table \ref{tab2}.

\begin{table*}
\small \centering \label{tab2} \caption{The Gaussian components
fitted with  MODELFIT in the Difmap package. Col. (1), component's
name. Col. (2), flux density of each component in mJy. Col. (3-4),
distance and position angle with respect to component A (or A+B+C at
low frequencies). Col. (5-7),
 major and minor axes and position angle (P. A.) of component. } \label{tab2}
\begin{tabular}{lrrrrrr}
\hline \hline
         &    &\multicolumn{2}{c}{Relative Position} &\multicolumn{3}{c}{Size of Gaussian component}\\
\cline{3-4} \cline{5-7}
Component&Flux& Radius&$\Theta$   & Major   & Minor  & P. A.   \\
      & (mJy) & (mas)  & ($^\circ$)& (mas)   & (mas)  &($^\circ)$        \\
\hline
      &\multicolumn{6}{l}{$\nu=1.438$ GHz}                              \\
A+B+C &777.2  $\pm$ 67.8 &  0.00              &  -----~~             &   1.98  &  1.39  & $-$74.05         \\
E+F   &9.1    $\pm$ 3.0  &  8.24  $\pm$ 0.21  & $-$134.19 $\pm$ 1.46 &   $<1$  &  $<1$  &  -----~~         \\
G     &23.6   $\pm$ 8.0 &  39.38 $\pm$ 2.25  & $-$109.70 $\pm$ 3.27 &   19.36 &  19.36 &  -----~~         \\
\hline
      &\multicolumn{6}{l}{$\nu=1.667$ GHz}                              \\
A+B+C &1010.3 $\pm$ 90.0 &  0.00             &  -----~~              &   2.20  &  1.64  &  0.33            \\
E+F   &11.4   $\pm$ 4.0  &  8.89  $\pm$ 0.12 & $-$129.67 $\pm$ 0.80  &   2.35  &  2.35  &  -----~~         \\
G     &24.5   $\pm$ 10.5 &  32.10 $\pm$ 1.83 & $-$107.63 $\pm$ 3.27  &   13.70 &  13.70 &  -----~~         \\
\hline
      &\multicolumn{6}{l}{$\nu=2.270$ GHz}                              \\
A+B+C &1346.1 $\pm$ 124.4 &  0.00              &  -----~~             &  2.08   & 1.56   & $-$8.65          \\
E+F   &66.0   $\pm$ 13.9  &  7.39  $\pm$ 0.37  & $-$126.85 $\pm$ 2.87 &   0.79  &  0.79  &  -----~~         \\
G     &11.5   $\pm$ 3.2   &  33.32 $\pm$ 4.79  & $-$106.64 $\pm$ 8.19 &   11.40 &  11.40 &  -----~~         \\
\hline
      &\multicolumn{6}{l}{$\nu=4.987$ GHz}                              \\
A     &1320.0 $\pm$ 122.6 &  0.00             &  -----~~              &   0.74  &  0.48  & $-$57.68         \\
B     &551.9  $\pm$ 62.9  &  1.08 $\pm$ 0.09  & $-$21.28  $\pm$ 4.97  &   1.36  &  0.94  &  17.90           \\
C     &258.8  $\pm$ 35.0  &  1.22 $\pm$ 0.11  & $-$167.18 $\pm$ 5.14  &   0.73  &  0.73  &  -----~~         \\
E     &95.2   $\pm$ 17.4  &  6.63 $\pm$ 0.15  & $-$128.91 $\pm$ 1.28  &   0.24  &  0.24  &  -----~~         \\
F     &116.9  $\pm$ 21.4  &  7.50 $\pm$ 0.17  & $-$119.49 $\pm$ 1.26  &   0.65  &  0.65  &  -----~~         \\
\hline
      &\multicolumn{6}{l}{$\nu=6.668$ GHz}                              \\
A+B+C &1979.7 $\pm$ 206.2 & 0.00             &  -----~~             &   1.44  &  0.60  &  $-$3.46        \\
E+F   &139.7  $\pm$ 34.8  & 6.78 $\pm$ 0.50  & $-$129.20 $\pm$ 4.19 &   1.34  &  0.35  &  $-$72.47        \\
\hline
      &\multicolumn{6}{l}{$\nu=8.420$ GHz}                              \\
A     &1046.9 $\pm$ 97.4 & 0.00             &  -----~~             &   0.68  &  0.31  &  $-$49.07        \\
B     &317.4  $\pm$ 45.5 & 0.92 $\pm$ 0.10  & $-$26.76  $\pm$ 5.95 &   1.32  &  0.89  &  $-$14.40        \\
C     &280.9  $\pm$ 35.4 & 1.00 $\pm$ 0.05  & $-$146.91 $\pm$ 3.09 &   0.60  &  0.33  &  8.45            \\
E     &84.9   $\pm$ 17.1 & 6.78 $\pm$ 0.11  & $-$127.04 $\pm$ 0.88 &   0.61  &  0.53  &  $-$5.36         \\
F     &57.7   $\pm$ 15.5 & 7.69 $\pm$ 0.17  & $-$117.68 $\pm$ 1.25 &   0.85  &  0.80  &  $-$85.70        \\
\hline
      &\multicolumn{6}{l}{$\nu=15.366$ GHz}                             \\
A     &546.3  $\pm$ 56.4  & 0.00            &  -----~~              &   0.57  &  0.40  &  $-$52.80        \\
B     &39.7   $\pm$ 8.1   & 1.27 $\pm$ 0.07 & $-$29.28  $\pm$ 3.12 &   0.69  &  0.37  &  33.08           \\
C     &124.4  $\pm$ 17.3  & 1.18 $\pm$ 0.04 & $-$145.63 $\pm$ 1.82 &   0.33  &  0.24  &  9.96            \\
E     &38.3   $\pm$ 10.5  & 6.79 $\pm$ 0.11 & $-$126.76 $\pm$ 0.93 &   0.58  &  0.48  &  $-$79.03        \\
F     &16.7   $\pm$ 6.9   & 7.64 $\pm$ 0.22 & $-$117.49 $\pm$ 1.62 &   0.92  &  0.61  &  $-$27.37        \\
\hline
%\hline
\end{tabular}
\end{table*}

Our 1.4-GHz image reveals a new component marked `G' in Fig. 1a,
which is a weak and extended component located at $\sim$ 30 mas and
the position angle $\sim$ $-110^\circ$. Component G also appears in
the 1.7-GHz image Fig. 1b. It is resolved at 2.3 GHz in Fig. 1d when
we use the full visibility data. However, it can be restored in Fig.
1c when we use the short-baseline ($< 20$ M$\lambda$) data. The
consistence of the position of component G at the three frequencies
completely confirms the existence of component G.

At the higher ($> 2.3$ GHz) frequencies, we do not detect component
G after trying the different short-baseline data and different
weighting methods. It indicates that component G is intrinsically
weak. The weakness is also consistent with the extrapolation from
the decreasing spectrum between 1.7 and 2.3 GHz. The  radio
structure of OQ 208 consists of two mini-lobes (NE and SW) at 2.3
and  6.7 GHz. The NE lobe can be fitted with three components (A, B
and C) and the SW lobe with two components (E and F) at 5.0, 8.4 and
15 GHz. The 5.0 and 8.4 GHz images are consistent with the previous
observations (e.g. Stanghellini et al. \cite{sta97}; Fey et al.
\cite{fey96}). Kellermann et al. (1998, observed in 1995)  found
that there is a weak emission region including the core between the
two lobes at 15 GHz.  Comparing with their 15 GHz images, we did not
detect any weak ($>$ 1 mJy/beam) structure at 8.4 and 15.4 GHz
between the two lobes although we have tried different weighting
methods and (u, v) taper. It is too weak and most likely that the
central core region was in a relatively quiescent period at the
epoch of our observations. From the VLBA 2cm survey
\footnote{http://www.cv.nrao.edu/2cmsurvey/}, we see that the
emission from the central region is truly  detected with the
observations on Apr. 7, Dec. 15, 1995, Apr. 22, May 16, Oct. 27,
1996, Aug. 28, 1997 and Mar. 7, 1998. But from the observations on
Oct. 30, 1998 to Apr. 28, 2006, nearly nothing is found in the
central region. This further confirms our results at 15.4 GHz.

The integrated flux densities of the two lobes, component G and the
whole source at each frequency are summarized in Table 3, and their
errors are within 10$\%$ of their corresponding data. From the
literature (Stanghellini et al. \cite{sta97}, \cite{sta05}) and our
data, we can see the total flux densities of OQ 208 at 1.4 and 1.7
GHz are stable from 1980. Our VLBI flux density of OQ 208 at 2.3 GHz
is 86$\%$ of the result observed with the telescope RATAN 600 at the
same frequency (Stanghellini et al. \cite{sta98}), indicating a
small decrease.

At 5.0 GHz, the total flux density of OQ 208 observed with VLA by
Dallacasa et al. (2000) decreased by about 10$\%$ compared with the
data given by Stanghellini et al. (1998). Comparing with the VLBI
observation by Stanghellini et al. (1997), our flux density of the
NE lobe decreased by $9\%$, while the SW lobe is stable. We note
that the total flux densities of OQ 208 observed with VLA are stable
within $5\%$ from Stanghellini et al. (1998) to Tinti et al. (2005)
at 8.4 GHz. The same is to the NE and SW lobes when we compare the
VLBI data of Fey et al. (1996) with ours.

\begin{table*}
\centering \label{tab3} \caption{The flux densities of OQ 208 in the
multi-frequency VLBI observations. } \scriptsize \label{tab3}
\begin{tabular}{rrrrrrrr}
\hline \hline
Freq. (GHz) & 1.438             & 1.667             & 2.270              & 4.987              & 6.668              & 8.420              & 15.365  \\
\hline
Total (mJy) & 809.9 $\pm$ 72.6  & 1046.2 $\pm$ 95.5 & 1423.6 $\pm$ 130.7 & 2342.8 $\pm$ 205.9 & 2119.4 $\pm$ 217.4 & 1787.8 $\pm$ 159.9 & 765.4 $\pm$ 73.6\\
NE (mJy)    & 777.2 $\pm$ 67.8  & 1010.3 $\pm$ 90.0 & 1346.1 $\pm$ 124.4 & 2130.7 $\pm$ 189.0 & 1979.7 $\pm$ 206.2 & 1645.2 $\pm$ 148.2 & 710.4 $\pm$ 68.9 \\
SW (mJy)    & 9.1   $\pm$ 3.0   &  11.4  $\pm$ 4.0  & 66.0   $\pm$ 13.9  & 212.1  $\pm$ 30.1  & 139.7  $\pm$ 34.8  & 142.6  $\pm$ 24.4  & 55.0  $\pm$ 14.8\\
G (mJy)     & 23.6  $\pm$ 8.0   & 24.5   $\pm$ 10.5  & 11.5   $\pm$ 3.2   & -~-~-              & -~-~-              & -~-~-              & -~-~- \\
\hline
%\hline

 \end{tabular}
\end{table*}

The total flux density of OQ 208 observed with VLA at 15.4 GHz is
1302 mJy (Dallacasa et al. \cite{dal00}) and 1139 mJy (observed in
2002, Tinti et al. \cite{tin05}). We have plotted the VLBA total
flux densities of OQ 208 at 15.4 GHz from the VLBA 2cm survey in
Fig. 2. From the figure, we find that its VLBA flux density is
decreasing from the epoch 1999.55. This may result from the fading
of the core since that time as we noted. Moreover, we find the flux
density of the NE lobe decreased by $37\%$ (reprocessed by us) from
May 5, 2001 to Apr. 28, 2006, while the SW lobe shows stable. This
also accounts for the drop of the VLBA total flux density of OQ 208
at 15.4 GHz.

\begin{figure*}
\centering \label{fig2}
\includegraphics[width=0.6\textwidth]{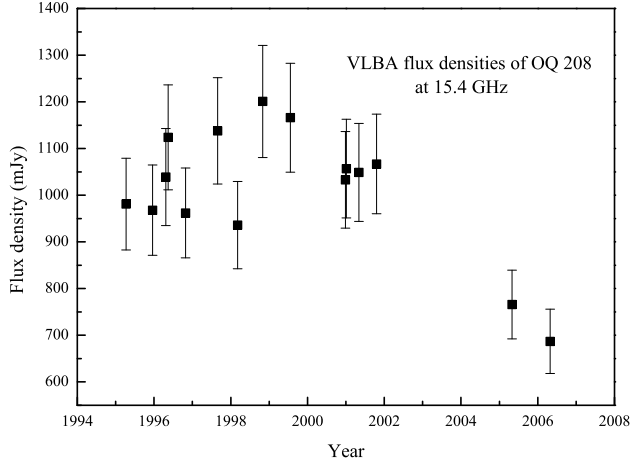}
\caption{The VLBA total flux densities of OQ 208 at 15.4 GHz. Data
are from the VLBA 2cm survey with 10$\%$ errors added, and the data
on the epoch of 2005.33 is ours.} \label{fig2}
\end{figure*}

We think that the core of OQ 208 was flaring in 1995-1996, or even
more earlier, and gradually fading out afterwards. The fading of the
NE lobe at 15.4 GHz is likely due to less energy supply from the
relatively quiescent central engine in recent years. With this
explanation, drop of the flux densities of OQ 208 at lower
frequencies should be expected to occur with some time delay.

\subsection{Proper motion}
We also estimated the separation of components A and E using the
available 8.4 GHz VLBI measurements over a period of about 130
months. Data for epochs 1994.52, 1997.08, 1997.25, 1997.38, 1997.56,
1997.96, 1998.48, 2002.04 and 2005.33 are from Liu et al. (2000),
Wang et al. (2003), the radio reference frame image database
(RRFID)\footnote{http://rorf.usno.navy.mil/rrfid.shtml} and this
paper. The estimate of the position errors is obtained by using the
standard deviation between the positions found by the different
tasks, e.g. the tasks IMFIT, OMFIT, JMFIT and SLIME in AIPS, and
MODELFIT in DIFMAP for the same components except ours that is
discussed in section 2. Typical values for the position errors are
between 0.03 and 0.2 mas, depending on the amount of uv-data. We
plot the separation between components A and E as a function of the
observing epoch in Fig. 3, and perform a linear fit to these data to
estimate the separation rate between them, which is 0.031 $\pm$
0.006 mas yr$^{-1}$. This proper motion has confirmed the previous
estimate of 0.033 $\pm$ 0.013 mas yr$^{-1}$ by Stanghellini et al.
(2000) with longer timescale data. The redshift of OQ 208 locates
the source at a distance of $\sim$ 350~Mpc, at which an angular size
of 1 mas corresponds to a linear size of about 1.46 pc. Then we
obtain a projected jet speed of 0.074 $\pm$ 0.014 $c$. Assuming an
inclination of 45$^\circ$ between the jet and the line of sight
(Stanghellini et al. \cite{sta97}), we get the actual jet velocity
of 0.105 $\pm$ 0.020 $c$ and obtain a kinematic age of 219 $\pm$ 42
yr for the radio source.

We also used the data from the VLBA 2cm survey on April 7, 1995 and
ours at 15.4 GHz to estimate the separation of components A and E.
The result is 0.028 $\pm$ 0.012 mas yr$^{-1}$. Both results at 8.4
and 15.4 GHz indicate that OQ 208 is expanding indeed.

\begin{figure*}
\centering \label{fig3}
\includegraphics[width=0.6\textwidth]{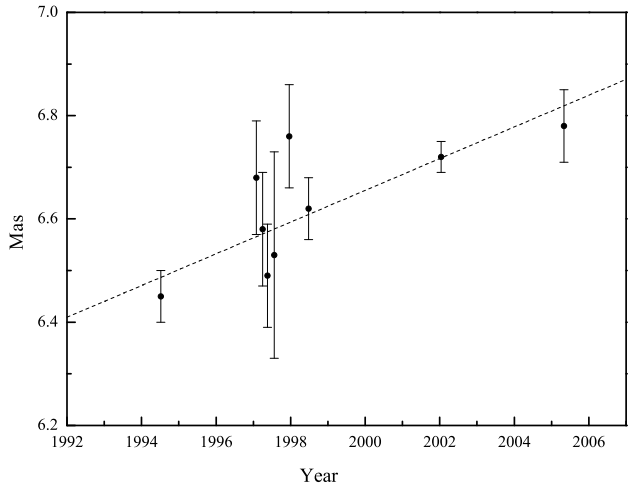}
\caption{Separation between components A and E in OQ 208 at 8.4 GHz.
} \label{fig3}
\end{figure*}

\section{Discussion}

\subsection{Property of Component G}

In the images at 1.4, 1.7 and 2.3 GHz (Fig. 1), component `G' is far
from the major components. It is difficult to explain the component
G in the scenario of CSO. Component G could be a remnant emission
from a previous radio burst, and  OQ 208 may be a recurrent radio
source. The steep spectral index ($\alpha = -2.4$) between 1.7 and
2.3 GHz further supports the remnant explanation for its emission
losses (Marecki et al. \cite{mar03}). Assuming a similar proper
motion in the recurrent scenario, we could estimate that the age of
component G is at least 1000 yr, which is much older than the CSO's
age of $\sim$ 220 yr.

\begin{figure*}
\label{fig3} \centering
\includegraphics[width=0.6\textwidth]{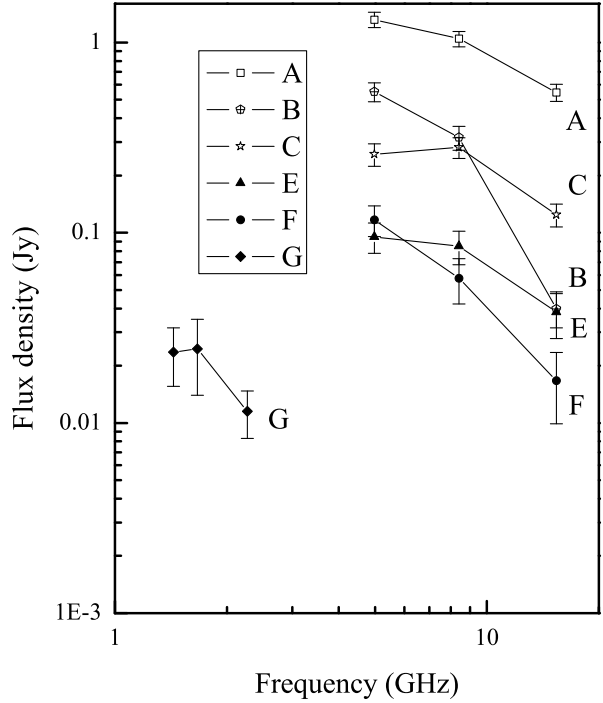}
\caption{Spectra of components A, B, C, E, F and G in OQ208.}
\label{fig3}
\end{figure*}

With the same resolution, the  flux densities of the component G are
$S_\mathrm{2.3 GHz}$ = 11.5 mJy, $S_\mathrm{1.7 GHz} = 24.5$ mJy and
$S_\mathrm{1.4 GHz}$ = 23.6 mJy. It  has an inverted spectrum with
the turnover frequency ($\leq 1.7$ GHz) lower than other components
shown in Fig.~4. If the turnover of G component is mainly caused by
SSA, then the turnover frequency in a homogenous, incoherent
synchrotron radio source with a power-law electron energy
distribution is given by Kellermann \& Pauliny-Toth (\cite{kel81})
\begin{equation}\label{eq1}
    \nu_\mathrm{t} \approx 8 B^{1/5}S_\mathrm{t}^{2/5} \theta^{-4/5}(1+z)^{1/5}~~~~\mathrm{GHz}
\end{equation}
where B is the magnetic field in Gauss, $S_\mathrm{t}$ the flux
density at the peak in Jy, $\theta$ the angular size in mas and $z$
the redshift. If we take $\nu_\mathrm{t} = 1.7$ GHz,  $S_\mathrm{t}
= 0.02$ Jy and $\theta = 13.7$ mas, the estimated magnetic field is
B $\sim 3.5\times10^4$ Gauss, too large to be real. The estimate
indicates that the SSA does not take effect at frequency $>$ 1.4
GHz. Therefore, FFA could be responsible for its spectral turnover.

\subsection{Absorption mechanisms}

It can be estimated from Table 3 that the two lobes in OQ 208 are
highly asymmetric with a flux density ratio between the north-east
(NE) lobe and south-west (SW) lobe of 85 $\pm$ 21 and 89 $\pm$ 23 at
1.4$-$1.7 GHz, much larger than the ratios of 20 $\pm$ 2, 10 $\pm$
1, 14 $\pm$ 2, 12 $\pm$ 1 and 13 $\pm$ 2 at frequencies 2.3, 5.0,
6.7, 8.4 and 15.4 GHz, respectively. The asymmetric free-free
absorption has been used to interpret the large flux density ratio
between NE and SW lobes (Kameno et al. 2000). Considering the radio
photons from the receding lobe is more scattered than that from the
approaching lobe in the line of sight, Thomson scattering was also
introduced to the flux density ratio of OQ 208, but it only
contributes a factor of 1.7 for the ratio (Liu et al. \cite{liu03}).
Xie et al. (2005) introduced the Doppler effect into the fit to the
observed spectra, and found that two models, i.e. the FFA+beaming
model and the SSA+beaming model, can fit the spectrum of the NE lobe
equally well. Our observations offer the wider spectral measurements
by adding a new low frequency 1.4 GHz data, and thus can be used to
test the different models. However, it is still hard to discriminate
which absorption mechanism is the dominant process for the NE lobe.
As for the SW lobe that has a turnover frequency $\sim$ 4 GHz,  SSA
only is definitely not enough to account for a steep spectral index
of $\alpha = 5.7$ between 1.7 and 2.3 GHz because the value is much
larger than the maximum attainable spectral index ($\alpha = 2.5$)
by the SSA model.

Fig. 4 shows the spectra of the pc-scale components of OQ 208. The
inner components A, C and E have a similar spectrum. Compared with
the inner components, the outer components B and F have steeper
high-frequency spectra. The flux density ratio between the counter
components E and F is 2.3 at 15 GHz, 1.5 at 8.4 GHz and 0.8 at 5
GHz. The decrease at 5 GHz could be further explained by the small
free-free opacity difference in the SW lobe. Using the uniform
FFA-opacity model (e. g. Kameno et al. \cite{kam01}), component E
has a turnover frequency $\nu_\mathrm{t} = 5.6$ GHz and component F
has $\nu_\mathrm{t} = 3.7$ GHz. Based on the projected distance of
component G from the core, we can give a lower limit to the
characteristic size of the external plasma, i.e. $\sim 57$ pc, which
is larger than that of any components. Therefore, the assumption of
the uniform FFA-opacity model is effective for the two components.
Both the decrease of the flux ratio and the estimated turnover
frequencies agree well, indicating that component E has a higher FFA
opacity than component F at the same emission frequency.
Furthermore, component G has the lower turnover frequency ($\leq
1.7$ GHz) than components E and F. Therefore, it seems that the
turnover frequency decreases from the inner component E to the outer
component G. This infers that the FFA is stronger in the inner
region.
\subsection{Free-free radiation}
Free-free radiation is thermal radiation. The radiation from the
source propagates through the foreground plasma, whose excitation
temperature is the same as the electron temperature $T_e$, and
attenuated. The spontaneous thermal radiation from the plasma is
added. Thus, the observed brightness temperature $T_b$ can be
written as:
\begin{equation}\label{e6}
    T_b = T_{b0}\exp (-\tau_\mathrm{f} \nu^{-2.1}) +
          T_e [1-\exp(-\tau_\mathrm{f} \nu^{-2.1})]~,
\end{equation}
where $T_{b0}$ is the intrinsic brightness temperature of radio
source in K and $\tau_\mathrm{f}$ the FFA coefficient at 1 GHz. If
the source is optically thin ($\nu > \nu_\mathrm{t}$), then $T_b
\approx T_{b0}$. For OQ 208, the brightness temperature of the lobe
or hot spot is $\sim 10^{9}$ K (e.g. Liu et al. \cite{liu02}), which
is much higher than the thermal temperature $T_e \sim 10^5$ K
(Kameno et al. \cite{kam01}). If the source is optically thick ($\nu
< \nu_\mathrm{t}$), the second item will be close to $T_e$. We will
expect that $T_{b0}$ increases with the decrease of frequency and is
close to the inverse Compton limit $\sim10^{12}$ K at a certain
frequency if the source has a power-law spectrum with $\alpha < 0$.
If we adopt $\tau_f = 8$ (Xie et al. \cite{xie05}), $\exp(-\tau)
\sim 3\times10^{-4}$ at 1 GHz and the first item is still much
larger than $T_e$. Therefore, for our observations at frequencies
greater than 1~GHz, the free-free radiation can be omitted.

\section{Summary and Conclusions}
We present the results of the quasi-simultaneous VLBI observations
of the GPS radio source OQ 208 at seven frequencies (1.4, 1.7, 2.3,
5.0, 6.7, 8.4 and 15.4 GHz). We detected a weak and extended
component G at about 30 mas in the position angle of $-110^\circ$ at
1.4, 1.7 and 2.3 GHz. The component G is supposed to be a relic
emission from an old radio burst and its age is estimated to be at
least 1000 yr assuming a similar jet proper motion in the CSO.
Though both SSA and FFA can fit the spectrum of the NE lobe well,
the SW lobe must undergo FFA considering its rising spectral index
of $\alpha = 5.7$ between 1.7 and 2.3 GHz, larger than 2.5 the
maximum attainable spectral index for SSA. The FFA is stronger in
the inner region than in the outer region.

We estimate a separation speed of 0.031 $\pm$ 0.006~mas/yr between
components A and E based on the 8.4 GHz VLBI observations at 9
epochs and further estimate a kinematic age of 219 $\pm$ 42 yr for
OQ 208. The proper motion (0.105 $\pm$ 0.020 $c$) of the lobe
confirms that estimated by Stanghellini et al. (2000) with longer
timescale data.

\begin{acknowledgements}

We thank the referee for valuable comments and J. Wrobel for
providing us the 1.4 GHz VLBA data. VLBA (Very Long Baseline Array)
is a facility of the National Radio Astronomy Observatory (NRAO),
operated by Associated Universities, under cooperative agreement
with the National Science Foundation. We thank the MOJAVE (Lister
and Homan, 2005, AJ, 130, 1418) and 2cm Survey (Kellermann et al.,
2004, ApJ, 609, 539) programs. This work has made use of Radio
Reference Frame Image Database of the United States Naval
Observatory. The European VLBI Network is a joint facility of
European, Chinese, South African and other radio astronomy
institutes funded by their national research councils. Z.-Q. Shen
acknowledges the support in part by the National Natural Science
Foundation of China (grant 10573029), Program of Shanghai Subject
Chief Scientist (06XD14024), and the One-Hundred-Talent Program of
Chinese Academy of Sciences.
\end{acknowledgements}

\end{document}